\documentclass[12pt,preprint]{aastex}

\slugcomment{Submitted to the Astrophysical Journal}
\shorttitle{Dust Properties of TMC-1C}
\shortauthors{Schnee, S. L. & Goodman, A. A.}

\begin{document}

\newcommand{\nthp}{N$_2$H$^+$(1-0)}
\newcommand{\kms}{km s$^{-1}$}

\title{Density and Temperature Structure of TMC-1C from 450 {\it and}
       850 \micron\ Maps}
\author{S. Schnee and A. Goodman}
\affil{Harvard-Smithsonian Center for Astrophysics, 60 Garden
       Street, Cambridge, MA 02138}
\email{sschnee@cfa.harvard.edu}

\begin{abstract}
	We have mapped the central 10\arcmin$\times$10\arcmin\ of the
dense core TMC-1C at 450 and 850 \micron\ using SCUBA on the James
Clerk Maxwell Telescope.  The unusally high quality of the 450
\micron\ map allows us to make a detailed analysis of the temperature
and column density profiles of the core.  We find that the dust
temperature at the center of TMC-1C is $\sim$ 7 K, rising to $\sim$ 11
K at the edges.  We discuss the possibility and effects of a variable
emissivity spectral index on the derived mass profile.  The low dust
temperature of TMC-1C results in a high derived mass for the core,
significantly larger than the virial mass estimated from the linewidth
of the \nthp\ transition.  This result is valid within a wide range of
dust properties and ellipticities of the core.  The \nthp\ spectra,
taken with the IRAM 30m telescope, show signs of self-absorption,
which provide evidence of sub-sonic infall motions.  The derived
density profile and infall velocity is compared to the predictions of
several popular star formation models, and the Bonnor-Ebert model is
the best fit analytic model.

\end{abstract}


\section{Introduction}
	Prestellar cores are self-gravitating condensations in
molecular clouds that are much denser and colder than their
surrounding medium.  These objects (such as L1544 and L1544F) are
believed to be on the verge of collapse, and infall motions are not
uncommon \citep{Williams99b, Crapsi04}.  Because prestellar cores are
at a critical point on the way to becoming a star, their properties,
such as density and temperature distribution, as well as their
velocity field, are of particular interest.  The density and
temperature profiles for prestellar cores are difficult to obtain from
molecular line data, because gases can deplete onto the dust grains at
the high ($n\gtrsim10^{4}$ cm$^{-3}$) densities and low temperatures
($T < 10$ K) present in the interiors of cores.  A more accurate
estimate can be obtained from measurements of the dust emission, which
peaks at submillimeter wavelengths and is often optically thin.
Observations at multiple submillimeter wavelengths can be used to
determine such properties as the temperature, density, emissivity, and
spectral index of the dust \citep{Hildebrand83}.  Kinematic
information, such as infall, outflow, and rotation, can be determined
by molecular line observations of carefully chosen high and low
density tracers \citep{Caselli02b,Belloche02,DiFrancesco01}.

	TMC-1C is a starless core in the Taurus molecular cloud
complex, at a distance of 140 pc \citep{Kenyon94}.  Taurus is known to
be a site of low mass, somewhat isolated star formation.  TMC-1C has
previously been observed to be a ``coherent core,'' meaning that its
velocity dispersion is constant, at slightly more than the sound
speed, over a radius of 0.1 pc \citep{Barranco98, Goodman98}.  In
addition, it shows a velocity gradient consistent with solid body
rotation, at 0.3 km s$^{-1}$ pc$^{-1}$ \citep{Goodman93}.  Because
TMC-1C is fairly nearby and has interesting kinematics, it is a good
candidate starless core for further study.

	In this paper we use continuum data taken with SCUBA at 450
and 850 \micron\ to determine the temperature and mass distribution of
the dust in TMC-1C, and spectral line maps from the IRAM 30m dish to
provide velocity information.  In Section \ref{EMISSIVITY} we note how
changes in the assumed dust emissivity alter the conclusions of our
analysis.  In Section \ref{UNIFORM} we compare the mass derived from
our multiwavelength observations to the mass we would have derived if
only 850 \micron\ data had been available to us.  In Section
\ref{COMPARISON}, various collapse models for star formation are
considered, and their predictions for infall speed and density profile
are compared to observations. Finally, we conclude that for a wide
range of assumptions, TMC-1C is a collapsing coherent core that is not
especially well fit by any of the star formation models that we
available.

\section{Observations}
\subsection{Continuum} \label{SCUBAOBS}

	We observed a 10\arcmin$\times$10\arcmin\ region of TMC-1C in
Taurus with SCUBA \citep{Holland99} on the JCMT in exceptionally
stable grade 1 weather.  We used the standard scan-mapping mode,
creating 850 \micron\ and 450 \micron\ maps simultaneously
\citep{Pierce-Price00, Bianchi00}.  Three chop throw lengths of
30\arcsec, 44\arcsec, and 68\arcsec\ were used in both the right
ascension and decination directions.  The JCMT has beam widths of
7.5\arcsec\ at 450 \micron\ and 14\arcsec\ at 850 \micron\ which
subtend diameters of 0.005 and 0.01 pc, respectively, at the distance
of Taurus.  Pointing during the observations was typically good to
3\arcsec\ or better.

\begin{figure}
\centerline{\includegraphics[width=3.5in]{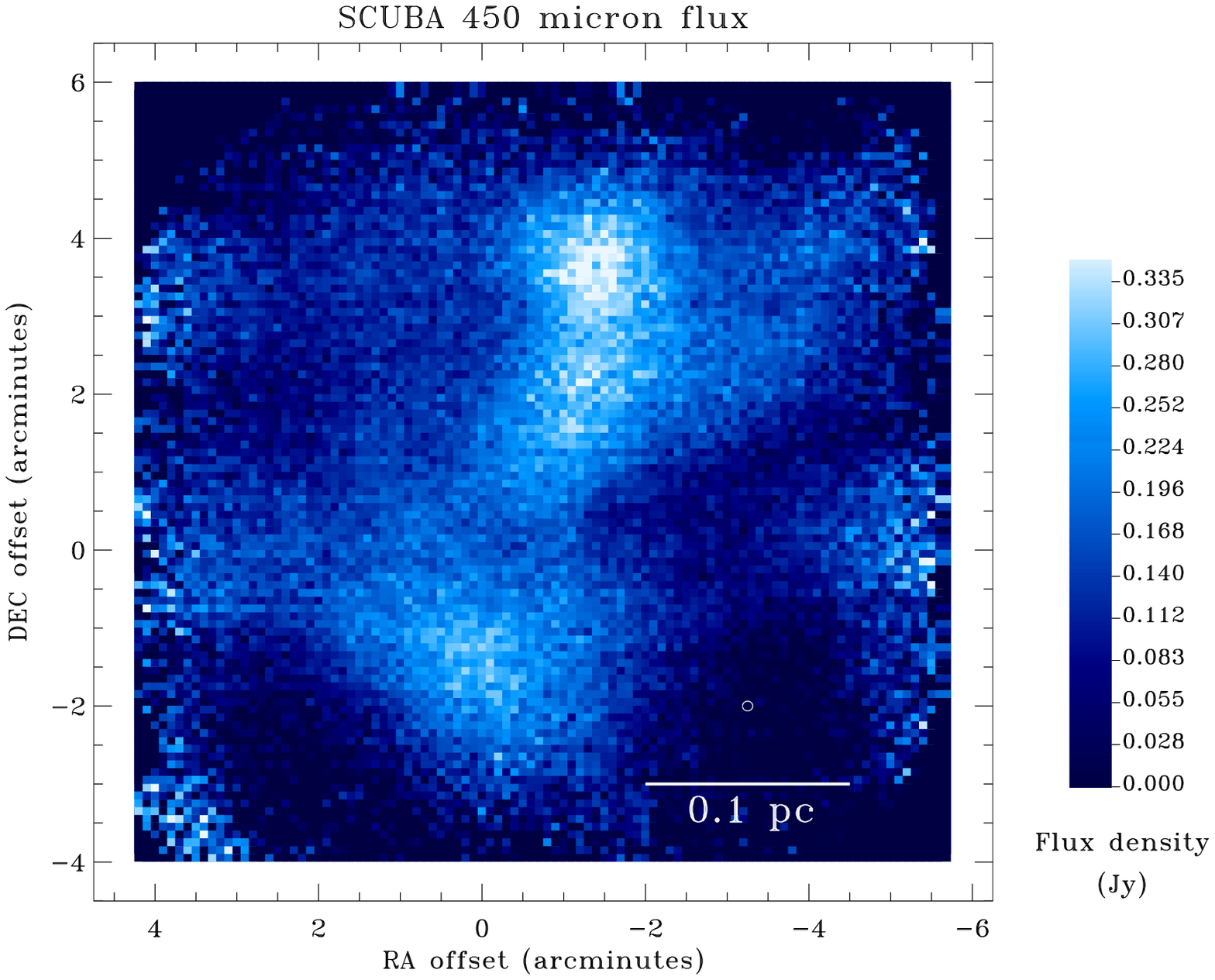}}
\centerline{\includegraphics[width=3.5in]{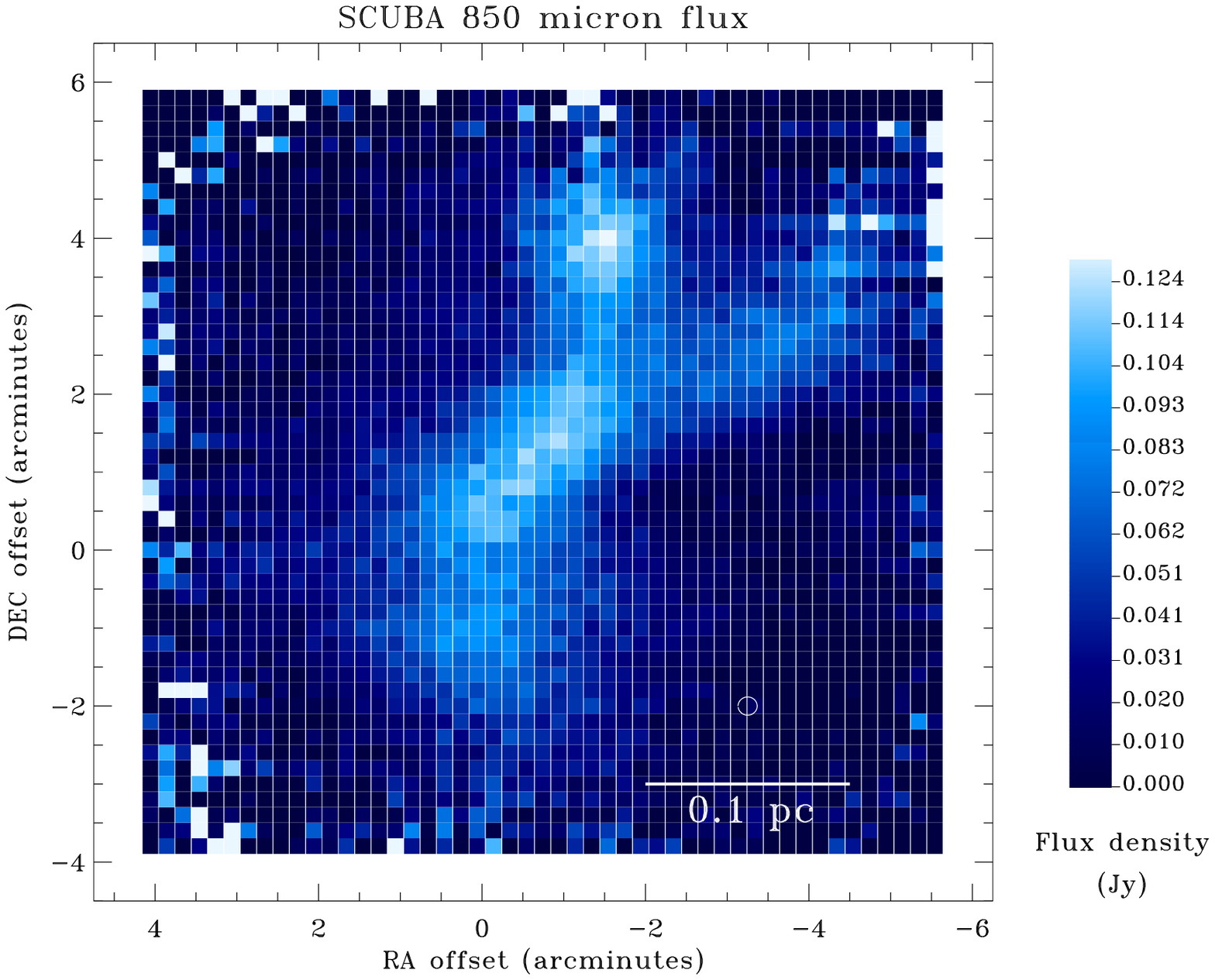}}
\figcaption{The 450\micron\ (top) and 850\micron\ (bottom) emission 
	    maps of TMC-1C. The SCUBA beam size is displayed in each 
	    map.  The coordinates of the (0,0) position are RA=4:41:38.8 
	    DEC=+25:59:42 (J2000).  \label{EMISMAPS}}
\end{figure}

	The data were reduced using the SCUBA User Reduction Facility
(SURF) \citep{Jenness98}.  The data were flatfielded, extinction
corrected using skydips, despiked, baseline corrected, and had sky
noise removed.  Maps were then made using Emerson Fourier
deconvolution \citep{Emerson95}.  The resultant maps were calibrated
with the source CRL618, with uncertainties in the calibrator's flux
densities of $\sim$3.7\% at 850 \micron\ and $\sim$12.5\% at 450
\micron.  The noise level of the 850 \micron\ map is 0.009 Jy/beam,
and 0.053 Jy/beam in the 450 \micron\ map.  Not that the scan-mapping
mode is insensitive to structures on scales greater than a few times
the largest chop throw.  The flux maps can be seen in Figure
\ref{EMISMAPS}.
   
\subsection{Spectral Line} \label{IRAMOBS}	
\begin{figure}
\plotone{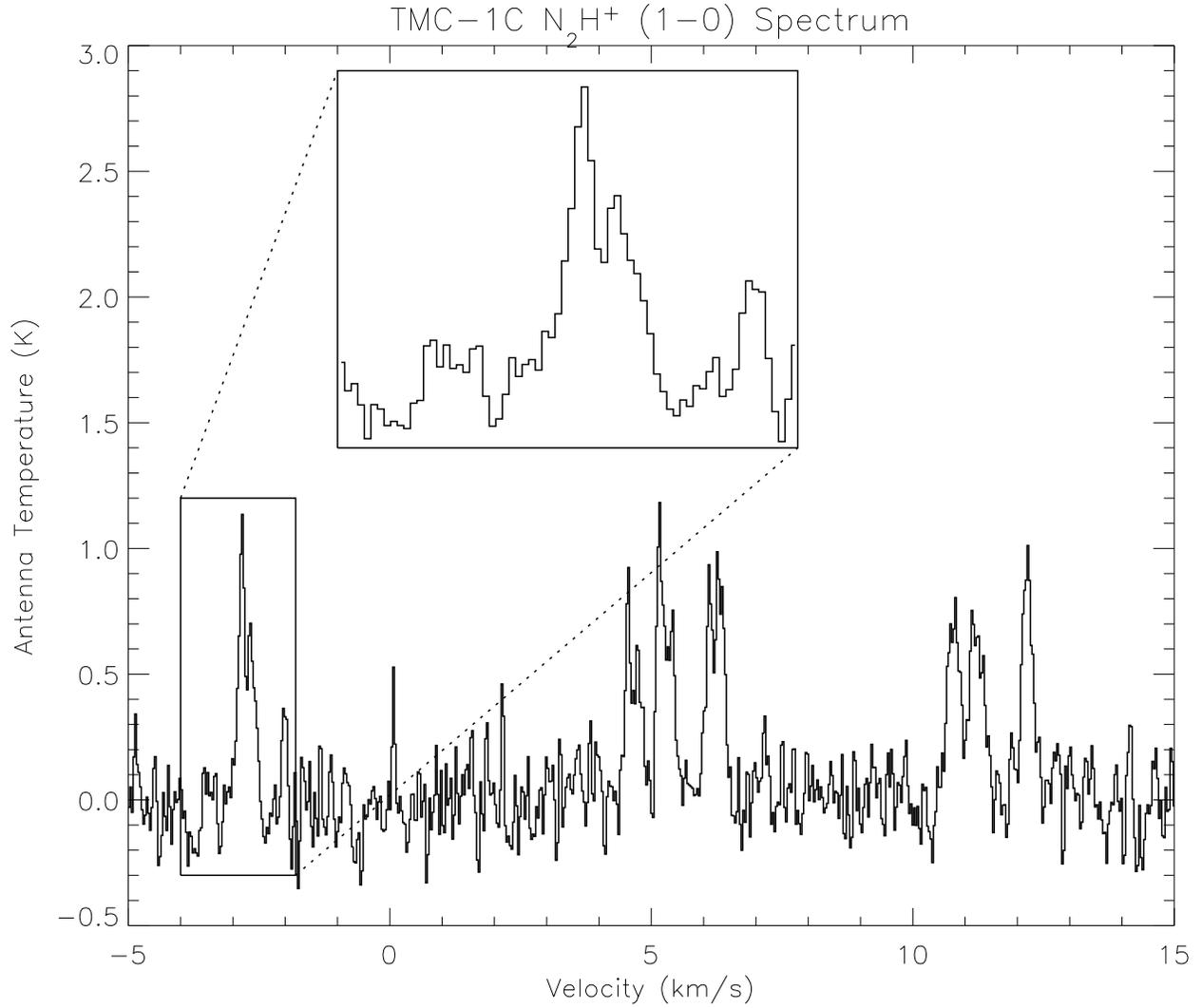}
\caption{\nthp\ spectrum from the (0,0) position.  Note the seven hyperfine
	 components.  The isolated component is shown at $\sim$ -2.8 \kms.
	 The velocity scale is only relative to the LSR for the central 
 	 component. \label{NTHPSPEC}}
\end{figure}

	The 3 mm \nthp\ observations shown in this paper were taken
with the IRAM 30m telescope, and will be discussed in more detail in
an upcoming paper \citep{Schnee05}.  A beam sampled map of the inner
$\sim$2' of TMC-1C was obtained in frequency switching mode with a
spectral resolution of 0.06 \kms. The data were reduced using the
CLASS package developed jointly between Observatoire de Grenoble and
IRAM.  Second order polynomial baselines were subtracted from the
data, and the seven hyperfine components of the \nthp\ spectra were
fit simultaneously with Gaussians for emission and absorption
\citep{Caselli95}.  The properties of the \nthp\ transition at the
central position are shown in Table \ref{NTHPTAB} and the spectrum at
that position is shown in Figure \ref{NTHPSPEC}.

\section{Dust Properties}
	Thermal emission from dust can be used to determine such
physical properties as temperature, density, and mass.  However, the
derivation of these quantities requires knowledge of the somewhat
poorly known emission characteristics of the dust grains.  In this
section we discuss our assumptions and methods to derive the
properties of TMC-1C from our sub-mm continuum observations.

\subsection{Assumed Parameters} \label{ASSUMED}
	The dust is assumed to emit as a modified blackbody, with
emissivity parameter $Q = (\lambda/\lambda_0)^{-\beta}$.  The flux
coming from the dust at a particular wavelength is therefore given by
\citep{Mitchell01}
\begin{equation} \label{DUSTFLUX}
S_{\lambda} = \Omega B_{\lambda}(T_d) \kappa_{_{\lambda}} \mu m_H N_H
\end{equation}
where
\begin{equation} \label{BLACKBODY}
B_\lambda(T) = \frac{2hc^2}{\lambda^5} \frac{1}{\exp(hc/\lambda kT)-1}
\end{equation}
and 
\begin{equation} \label{KAPPA}
\kappa_{_{\lambda}} = \kappa_{1300} \left(\frac{\lambda}{1300 \micron} 
	       \right)^{-\beta}
\end{equation}
In Equation \ref{DUSTFLUX}, $S_{\lambda}$ is the flux per beam;
$\Omega$ is the solid angle of the beam; $B_{\lambda}(T_d)$ is the
black body emission from the dust at temperature $T_d$; $\kappa_{1300}
= 0.005$ cm$^2$ g$^{-1}$ is the emissivity of the dust grains at 1300
\micron\ \citep{Andre96, Preibisch93}; $m_H$ is the mass of the
hydrogen atom; $\mu = 2.33$ is the mean molecular weight of
interstellar material in a molecular cloud, and $N_H$ is the column
density of hydrogen nuclei.  For a true blackbody $\beta = 0$; for
amorphous, layerlattice material $\beta \sim 1$; for metals and
crystalline dielectrics $\beta \sim 2$, cf \citep{Henning95}.  We
assume the emission is optically thin at both 850 and 450 \micron,
which is justified for H$_2$ column densities less than $\sim 10^{24}$
($\sim$ 500 magnitudes of visual extinction) \citep{Zucconi01}.

\subsection{Derivation of Temperature} \label{TEMPERATURE}
	Because the thermal emission from cold ($\sim$10 K) dust at
450 and 850 \micron\ is both optically thin and near the peak of the
modified blackbody spectrum, SCUBA is well-suited to observations of
starless cores.  A dust color temperature can be determined from the
ratio of fluxes at two wavelengths.  At temperatures near 10 K, the
blackbody spectrum peaks around 300 \micron, so for data taken at 450
and 850 \micron\ the Rayleigh-Jeans approximation to the blackbody law
cannot be used.  The flux ratio therefore depends on the assumed
emissivity spectral index and the temperature \citep{Kramer03} as:

\begin{equation} \label{TEMPEQ}
\frac{S_{450}}{S_{850}} = \left(\frac{850}{450}\right)^{3+\beta}
                          \frac{\exp(17\textrm{K}/T_{dust})-1}
			  {\exp(32\textrm{K}/T_{dust})-1}
\end{equation}  
where $hc$/$\lambda$$k$ is 17 K and 32 K at 850 and 450 \micron\ 
respectively.

	In order to derive the dust temperature in TMC-1C, we smoothed
both SCUBA maps to the same 14\arcsec\ resolution.  Because the 450
and 850 \micron\ SCUBA maps both have such high signal to noise, we
were able to determine the temperature independently at each
14\arcsec$\times$14\arcsec\ pixel and make a detailed temperature map,
assuming a constant value for $\beta$ throughout the
core\footnote{This restriction is relaxed in Section \ref{ESI}.}.  By
estimating a temperature at each point in TMC-1C from our
observations, we can calculate self-consistent profiles for column
density and mass.  A map of the derived dust temperature is shown in
Figure \ref{TEMPMAP}.  The derived dust temperatures range from 6 K to
15 K.

\subsection{Calculation of Extinction} \label{EXTINCTION}
	Given the dust color temperature at each position, the column
density can be calculated from the observed flux, and from this, the
equivalent visual extinction can be estimated.  Equation
\ref{DUSTFLUX} can be rearranged to determine the column density from
the observed flux and derived temperature.
\begin{equation}
N_H = \frac{S_{\lambda}}{\Omega B_{\lambda}(T_d) \kappa_{_{\lambda}} \mu m_H}
\end{equation}
The equivalent extinction is given by
\begin{equation}
A_V = N_H \frac{E(B-V)}{N_H} \frac{A_V}{E(B-V)}
\end{equation}
\begin{equation} \label{AVNH}
A_V = 5.3\times10^{-22} N_H
\end{equation}
where $N_H/E(B-V) = 5.8\times10^{21}$ cm$^{-2}$ is the conversion
between column density of hydrogen nuclei (for our assumed gas to dust
ratio) and the selective absorption, and $R_V = A_V/E(B-V) = 3.1$ is
the ratio of total-to-selective extinction \citep{Mathis90, Bohlin78}.
Note that this value of $R_V$ assumes a particular ``color'' for the
absorbing dust, which may be different from core to core, or even
within a single core.  The extinction map created using these
assumptions, and the value of $\kappa$ in Section \ref{ASSUMED}, is
shown in Figure \ref{EXTMAP}.  The derived extinction reaches a value
of $\sim$ 50 mag A$_V$ at the center of the map.  A map of the derived
column density is shown in Figure \ref{EXTMAP}.

\subsection{Derivation of the Emissivity Spectral Index} \label{ESI}
	Instead of assuming a constant value of $\beta$ to calculate
the temperature map as in Section \ref{TEMPERATURE}, one can
alternatively calculate the value of the emissivity spectral index at
each point in the map by choosing a constant value for the
temperature.  Using multi-transition spectral line data,
\citet{Tafalla02} find that the core L1544 has a constant {\it gas}
temperature of approximately 10 K (with spatial resolution of $\sim$
0.03 pc), so experimenting here with calculations involving a constant
{\it dust} temperature is justified.  Because the emissivity spectral
index appears in the equations used to derive the temperature and mass
of a core, knowledge of its variability within a core is clearly
valuable to the understanding of the physical properties of TMC-1C and
cores in general.  $\beta$ depends on the dust grain size
distribution, the composition of the mantle, and the surface area to
volume ratio of the dust particles, all of which can vary throughout a
core \citep{Ossenkopf94}.

	The derived emissivity spectral index map for $T = 10$ K at
all radii is shown in Figure \ref{BETAMAP}.  In the central 2
arcminutes ($\sim$0.08 pc) of the TMC-1C core, the value of $\beta$
takes values in the physically plausible range $0.5 \leq \beta \leq
2.0$.  To self-consistently determine the column density, temperature,
and emissivity spectral index of the dust simultaneously, we would
need at least one more continuum map at another wavelength.  For the
remainder of this paper we will assume that $\beta$ has a constant
value and that the dust temperature in TMC-1C is not fixed, unless
stated otherwise, although we understand that both quantities are
likely to vary throughout the core.
	
\subsection{Calculation of Mass} \label{MASS}
	For optically thin emission, Equation \ref{DUSTFLUX} can be 
rearranged to convert a measurement of dust emission to mass, so that   
\begin{equation}
M = \Omega \mu m_H N_H d^2 = \frac{S_{\lambda}d^2}{\kappa_{\lambda}B_{\lambda}
    (T_d)}
\end{equation}
where $M$ is the mass of an emitting volume and $d$ is the distance to it.
Using the conversions given above and assuming a distance to Taurus of 140 
pc, this equation is equivalent to 
\begin{equation} \label{MASS450}
M = 8.29 \times 10^{-3}\ S_{450} \left[\exp \left( \frac{32\ \textrm{K}}{T_d}
    \right) -1 \right] \left( \frac{\kappa_{450}}{0.026\ \textrm{cm}^2\ 
    \textrm{g}^{-1}} \right)^{-1} \left( \frac{d}{140\ \textrm{pc}} \right)^2
    \textrm{M}_{\odot}
\end{equation}
or
\begin{equation} \label{MASS850}
M = 0.145\ S_{850} \left[\exp \left( \frac{17\ \textrm{K}}{T_d} \right) -1 
    \right] \left( \frac{\kappa_{850}}{0.01\ \textrm{cm}^2\ \textrm{g}^{-1}} 
    \right)^{-1} \left( \frac{d}{140\ \textrm{pc}} \right)^2 
    \textrm{M}_{\odot}
\end{equation}

	The mass contained in each 14\arcsec$\times$14\arcsec\ pixel
is determined by Equations \ref{MASS450} and \ref{MASS850}.  Figure
\ref{VIRMAP} shows the mass profile calculated by adding the mass in
each pixel to annuli concentric around the peak column density.  The
mass in each pixel is determined using the 850 \micron\ flux in each
pixel and either the temperature assigned to each pixel shown in
Figure \ref{TEMPMAP} (the curve labeled ``Actual Temperature
Profile''), or a constant temperature assigned to every pixel.  The
total mass contained within a given radius of the TMC-1C column
density peak can be converted into a density profile, assuming a three
dimensional geometry, as discussed below.

\subsection{Temperature and Density Profiles} \label{PROFILES}
	Under the assumption of spherical symmetry, we can create
three dimensional temperature and density profiles from the 850 and
450 \micron\ maps.  To make the profiles, we break up the inner 2.2
arcminutes of TMC-1C into 14\arcsec\ wide concentric annuli, centered
on the position of the extinction peak.  The outermost annulus only
has flux contributed to it by the outermost spherical shell, and its
temperature and density can be calculated as in Sections
\ref{TEMPERATURE} and \ref{MASS}.  The next shell in recieves flux
from the two outermost spherical shells.  After subtracting the flux
contributed from the outermost shell, whose temperature and mass has
already been calculated, the remaining flux at 450 and 850 \micron\
comes from the second spherical shell, allowing its temperature and
density to be derived.  In this manner we are able to work our way
from the outside of TMC-1C in, finding the temperature and density of
each spherical shell.  The procedure is very similar to that described
in \citet{David01} for the Hydra A cluster of galaxies.

	In Section \ref{MASS}, the mass was calculated for each
element in the map independently of its neighbors by assuming that a
single temperature could be assigned to the entire column of material.
In this section, instead of dividing the central 2.2 arcminutes of
TMC-1C into isothermal cylinders, we break it up into isothermal
spherical shells.  The temperature and density profiles are fit by
broken power laws.  Each broken power law is described by five
parameters: two coefficients (for normalization), two exponents (the
slopes), and the break radius (where the two lines meet).  The
temperature and density profiles created in this way are shown in
Figure \ref{PROMAPS}.

	The derived temperature profile agrees well with the
theoretical temperature profile of an externally heated Bonnor Ebert
sphere presented in \citet{Goncalves04}, as well as the observed dust
temperature profile of L183 \citep{Pagani04}.  Outside the break
radius, the density profile of TMC-1C derived from the SCUBA data is
similar to density profiles derived from \nthp\ data and continuum
dust emission of other starless cores (Table \ref{DENSTAB}).  Inside
the break radius, TMC-1C has a considerably shallower density profile
than these other starless cores.  The starless cores to which we are
comparing TMC-1C did not have their density profiles determined with
the same concentric ring algorithm as used here, because
researchers lacked the data to make the detailed temperature map that
we were able to get from SCUBA.  The power law slopes of the density
profiles of these other cores can be thought of as isothermal
approximations to the ``true'' density profile.  For comparison, the
powerlaw fit to the density profile of TMC-1C under the assumption of
a constant $T_{dust} = 10$ K and using only the 850 \micron\ flux is
also shown in Figure \ref{CONSTPRO}, with the broken powerlaw fit
shown in Table \ref{DENSTAB}.

\section{Virial Mass and Infall} \label{VMI}
	The balance between outward turbulent and thermal pressures
and inward gravitational pull determines the stability of a dense
core.  In this section we derive the virial mass from \nthp\ data and
show that for a wide range of density profiles and shapes the virial
mass is significantly {\it lower} than the mass derived from dust
emission.  The \nthp\ emission peaks in the same place that the dust
column density peaks, so using this line to estimate the virial mass
is justified.  The \nthp\ self-absorption profile shows evidence of
infall motions, further suggesting the gravitational instability of
TMC-1C \citep{Schnee05}.

\subsection{Calculation of the Virial Mass} \label{VIRIAL}
	The virial mass of the inner core of TMC-1C can be estimated
from the observed line width of the isolated component of the \nthp\
spectra.  The total velocity dispersion has a non-thermal and a
thermal component, given by:
\begin{equation} 
\sigma_{TOT}^2 = \sigma_{NT}^2 + \sigma_T^2
\end{equation}

	The thermal velocity dispersion is given by the gas temperature by: 
\begin{equation} 
\sigma_T^2 = \frac{kT}{\mu_{tracer}}
\end{equation}

	The line width depends on the velocity dispersion as:
\begin{equation}
\Delta V = \sqrt{8\ln 2}\sigma_{TOT}
\end{equation}

Where $\sigma_{TOT}$ is the total velocity dispersion, $\sigma_{NT}$
is the non-thermal component of the velocity dispersion, $\sigma_T$ is
the thermal component of the velocity dispersion, $k$ is the Boltzmann
constant, $T$ is the gas temperature (assumed to be 10 K,
\citep{Tafalla02}), $\mu_{tracer}$ is the mass of the molecule
(N$_2$H$^+$ or H$_2$), and $\Delta V$ is the observed line width.
Note that the dust temperature is below 10 K throughout most of the
TMC-1C core, so the virial mass calculated for a 10 K gas is an
upper limit, if the gas and dust temperatures are coupled.

	The \nthp\ spectra were annularly averaged with bins of
20\arcsec, and the line width was then determined as a function of
distance from the core's center.  The \nthp\ line width is $\sim$ 0.25
$km$ $s^{-1}$, and is very nearly constant at all radii (as expected for
a coherent core).  The virial mass of an ellipsoid with a power law
density profile determined in this way is
\begin{equation} \label{MVIR}
M_{\rm vir} = \frac{5R}{G} 
              \frac{(\Delta v_{tot}(H_2))^2}{8 \ln 2} 
              \frac{1}{a_1 a_2}
\end{equation}
where $\Delta v_{tot}$ is the quadratic sum of the 10 K thermal line
width of H$_2$ and the non-thermal line width of \nthp, $a_1$ is a
parameter than depends on the density profile, and $a_2$ is a
parameter that depends on the ellipticity of the core.  The virial
mass is plotted as a function of radius in Figure \ref{VIRMAP}, with
the assumptions that the density profile is uniform ($a_1 = 1$) and
the core is spherical ($a_2 = 1$).  More sophisticated geometric
assumptions are discussed in Sections \ref{DENSITYDEP} and
\ref{SHAPEDEP}.

\subsubsection{Dependence on the Density Profile} \label{DENSITYDEP}
	We have shown in Section \ref{PROFILES} that the density in
TMC-1C is not uniform, and our estimate of the virial mass should take
this into account.  For a given line width, a more centrally condensed
cloud will have a lower virial mass than a uniform density cloud
\citep{Bertoldi92}.  This can be parameterized for a power-law density
distribution $\rho (r) \varpropto r^{-k}$
\begin{equation} \label{VIRDEN}
a_1 = \frac{(1 - k/3)}{(1 - 2k/5)}
\end{equation} \citep{Bertoldi92}.  

	The dependence of the virial mass on the density power law is
shown in Figure \ref{VIRPARM}.  For r$^{-1}$ and r$^{-2}$ density
distributions, the corrected virial masses are 90\% and 60\% of the
value calculated by Equation \ref{VIRDEN} for a uniform density
profile.  Thus, the steeper the density profile, the greater the
imbalance between the virial mass and the dust derived mass for
TMC-1C, making it more unstable.

\subsubsection{Dependence on the Shape of the Core} \label{SHAPEDEP}
	Our estimate of the virial mass also depends on the shape of
TMC-1C.  It has been shown that many clouds are elliptical in
projection, and usually prolate in three dimensions \citep{Myers91}.
Oblate clouds have slightly lower virial masses than a sphere whose
radius is determined by the plane-of-sky dimension and a prolate cloud
will have a larger virial mass than such a sphere.  If the axial
ratio, y, is less than 1 (an oblate cloud) then
\begin{equation} \label{VIRAX1}
a_2 = \frac{\arcsin(1-y^2)^{1/2}}{(1-y^2)^{1/2}}
\end{equation}
If the axial ratio, y, is greater than 1 (a prolate cloud) then
\begin{equation} \label{VIRAX2}
a_2 = \frac{\textrm{arcsinh}(y^2-1)^{1/2}}{(y^2-1)^{1/2}}
\end{equation} \citep{Bertoldi92}.  The dependence of the virial mass on 
the axial ratio is shown in Figure \ref{VIRPARM}.  For axial ratios
less than 10:1, the correction for the virial mass is less than a
factor of $\sim$ 3.  The virial mass as a function of shape and
density profile for TMC-1C are shown in Figure \ref{VIRDEP}, as are
the masses derived from the dust emission for three representative
values of $\beta$.  Notice that to make the virial mass equal to the
dust-emission implied mass at 0.06 pc in TMC-1C (see Figure
\ref{VIRMAP}), a prolate core would need to have an axial ratio of
20:1.  This is ruled out by our data, except for the very unlikely
case where we are viewing a cigar end-on.

\subsection{\nthp\ Self-Absorption} \label{N2H+}
	\nthp\ is typically used as an optically-thin, high-density,
low-depletion tracer of the kinematics of the interior of cores
\citep{Williams99, Caselli02a, Caselli02b}.  However, there is some
evidence of \nthp\ self-absorption in dense cores \citep{Caselli02b,
Williams99}.  In another paper we will show that \nthp\ is a better
match to the dust emission in TMC-1C than most carbon bearing
molecules, which tend to deplete onto dust \citep{Schnee05}.  In this
paper we show evidence of redshifted self-absorption in \nthp\ at
several positions near the peak of the dust emission.

	To determine if the splitting of the \nthp\ hyperfine
components is due to self-absorption or two superimposed velocity
components from different emission components along the same line of
sight, the seven hyperfine components can be compared with each other
to see if the location of the dip is constant amongst them.  If this
is the case, then kinematics are likely to be creating the
non-gaussian spectra, if not (with the shift increasing with the
statistical weight of the component), then self-absorption is probably
the cause.  The dip in the TMC-1C \nthp\ spectra does shift with
statistical weight in a way consistent with self-absorption, so we
believe that the double peaked spectra are not the result of two
kinematically distinct clouds.  While \nthp\ self-absorption is
uncommon enough to be interesting on its own, we use the \nthp\ data
in this paper to determine infall velocity of material close to the
center of the TMC-1C core.

	The isolated component of each \nthp\ spectrum (see Figure
\ref{NTHPSPEC}) was fit with two gaussians (one positive and one
negative) to find the approximate velocities for the emission and
absorption features, respectively.  If the interior of the core is
collapsing or flowing toward the center faster than the exterior, then
radiative transfer models show that the spectra should have two peaks,
with the blue peak higher than the red \citep{Zhou93, Myers96}.
Therefore, in an infalling region, the gaussian fit to emission will
be bluer than the gaussian fit to absorption.  In Figure
\ref{HISTOGRAM} we show the histogram of the difference between the
emission and absorption velocities for \nthp, where the absorption
velocities are nearly all redshifted relative to the emission
velocities, indicative of infall.  Using the simple two layer
radiative transfer model of a contracting cloud developed by
\citet{Myers96}, we find that the infall velocity averaged over the
central 0.05 pc of TMC-1C is 0.06 \kms.

\section{Discussion}
\subsection{450 and 850 \micron\ Emission Maps} \label{FLUXDIFF}
	The 450 and 850 \micron\ maps of TMC-1C look qualitatively
different in the lower half of the maps.  The 850 \micron\ map has a
peak at (-1',1'), though the 450 \micron\ map has no peak there.  The
450 \micron\ map peaks at (0',-1.5'), where the 850 \micron\ flux is
comparatively weak.  This suggests that the physical conditions in
TMC-1C are different from the pre-protostellar cores L1544 and L1698B,
in which the 450 and 850 \micron\ fluxes are well correlated
\citep{Shirley00}.

\subsection{Temperature Distribution} \label{TEMPDIST}
	In order to compare the observed dust temperature in a
starless core to theoretical models of externally heated cores, it is
necessary to have high quality maps in at least two wavelengths.  Our
SCUBA maps of TMC-1C provide some of the best available submillimeter
data of a starless core.  In Figure \ref{PROMAPS} we show that the
temperature profile is coldest at the center, with a temperature of
$\sim$ 6 K.  The temperature rises by $\sim$ 1 K out to 0.03 pc, after
which it rises more quickly, to a $\sim$ 12 K at a radius of 0.08 pc.

	Our observed dust temperture profile is a close match to that
shown in Figure 2 of \citet{Goncalves04}, in both temperature range
and shape of the profile.  Similarly, our dust temperature profile
also matches that in Figure 4b of \citet{Evans01}.  Both of these
models simulate the temperature profile of externally heated starless
cores with Bonnor-Ebert density distributions, which is a reasonable
approximation of TMC-1C.
	
\subsection{Dependence on $\beta$} \label{BETA}
	The parameters describing TMC-1C that we have derived above
all assume a constant emissivity spectral index throughout the cloud,
although a range of values of $\beta$ have been tested, between 1.0
and 2.0.  We have found that constant values of $\beta$ smaller than
1.0, the mass derived from the temperature and the 450 \micron\ flux
is not consistent with the mass derived from the temperature and the
850 \micron\ flux. In general, for fits with larger $\beta$, the
derived mass of the cloud increases and the temperature of the cloud
decreases.  For a low value of $\beta$ of 1.0, the virial mass for the
inner 0.06 pc of a uniform density sphere is still 2.5 times {\it
less} than the mass derived from dust emission (see Figure
\ref{VIRMAP}).  So, although the dust derived mass varies strongly
with $\beta$, the conclusion that the virial mass is smaller than the
dust derived mass of TMC-1C is robust (See Figure \ref{VIRDEP}).

\subsection{Dust Emissivity} \label{EMISSIVITY}
	The emissivity of dust has been measured with a combination of FIR 
flux maps and NIR extinction maps in the clouds IC 5146 and B68 
\citep{Kramer03, Bianchi03}.  Both studies find 850 \micron\ fluxes 
consistent with $\kappa_{850} = 0.01$ cm$^2$ g$^{-1}$, with uncertainties 
in the range of 30\% - 60\%.  TMC-1C is a starless core similar to B68 and 
those found in IC 5146, so we estimate that uncertainties in our 450 and 
850 \micron\ calibrations (see section \ref{SCUBAOBS}) and emissivities 
result in a derived mass that is accurate to within a factor of 2 for a 
given emissivity spectral index and dust to gas conversion.  The masses we 
derive from the dust emission are greater than the virial mass by more than 
a factor of two, so the uncertainty in the dust emissivity can not by 
itself change the conclusion that TMC-1C should be in a state of collapse.  
However, a perverse conspiratorial \emph{combination} of $\beta$, axial 
ratio, and $\kappa_{\nu}$ could still result in a dust mass lower than the 
\nthp\ derived virial mass.  Figure \ref{VIRDEP} shows contours of the 
virial mass as a function of ellipticity and density profile, and also plots 
the mass derived from the 450 and 850 \micron\ fluxes for representative 
values of $\beta$.  It is shown that for any density profile and any value 
of $\beta$ tested, TMC-1C is gravitationally unstable for a wide range of 
axial ratios.  If TMC-1C has a nearly spherical geometry, then the dust 
emissivity we use would have to be low by an unlikely factor of $\sim$ 3-5 
to make the virial mass and the dust derived mass approximately equal.	    

\subsection{Emissivity Spectral Index} \label{DIS_BETA}
\begin{figure}
\plotone{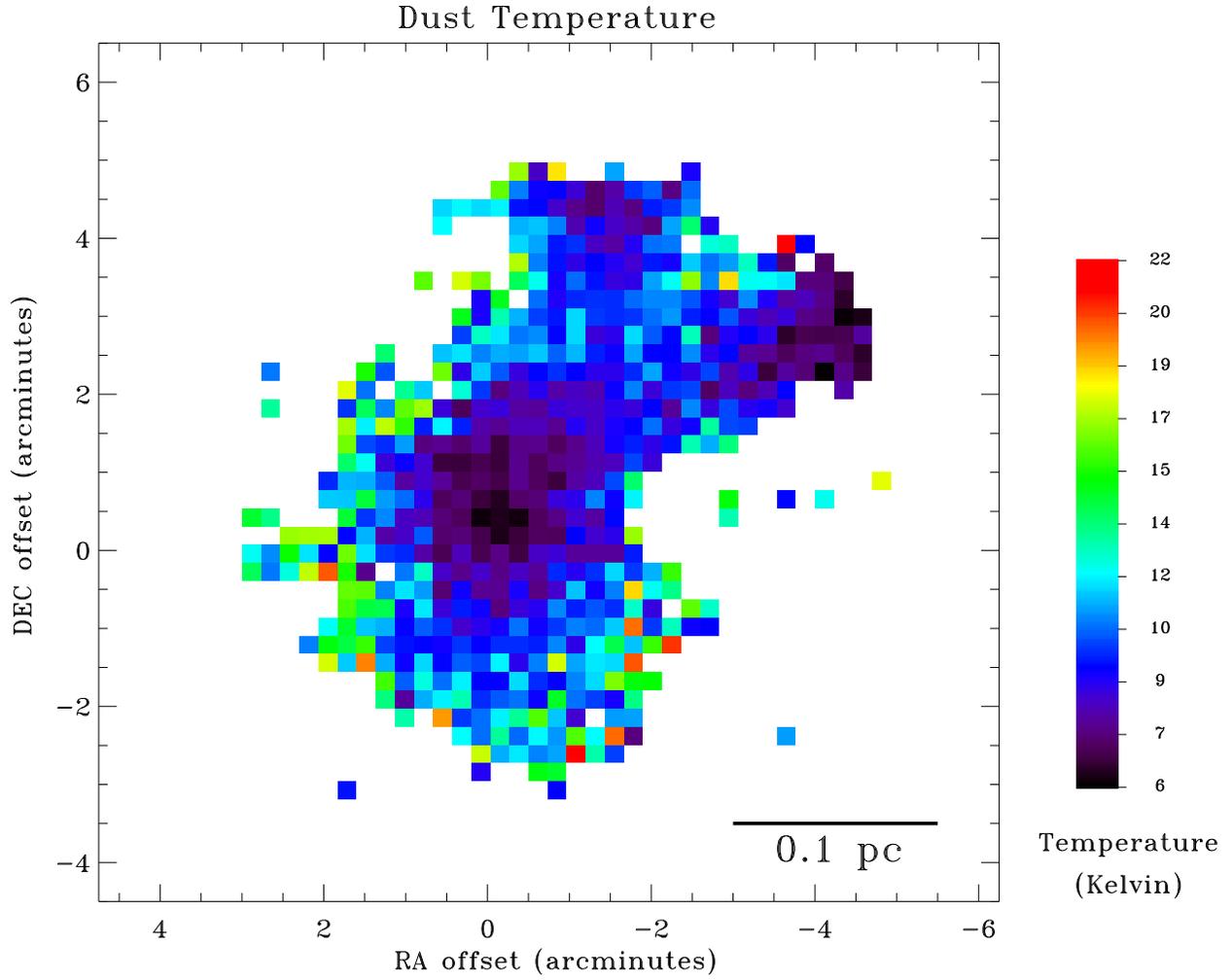}
\caption{Dust temperature derived from the 450 and 850\micron\ SCUBA maps
	with the choice of $\beta = 1.5$. \label{TEMPMAP}}
\end{figure}

\begin{figure}
\plotone{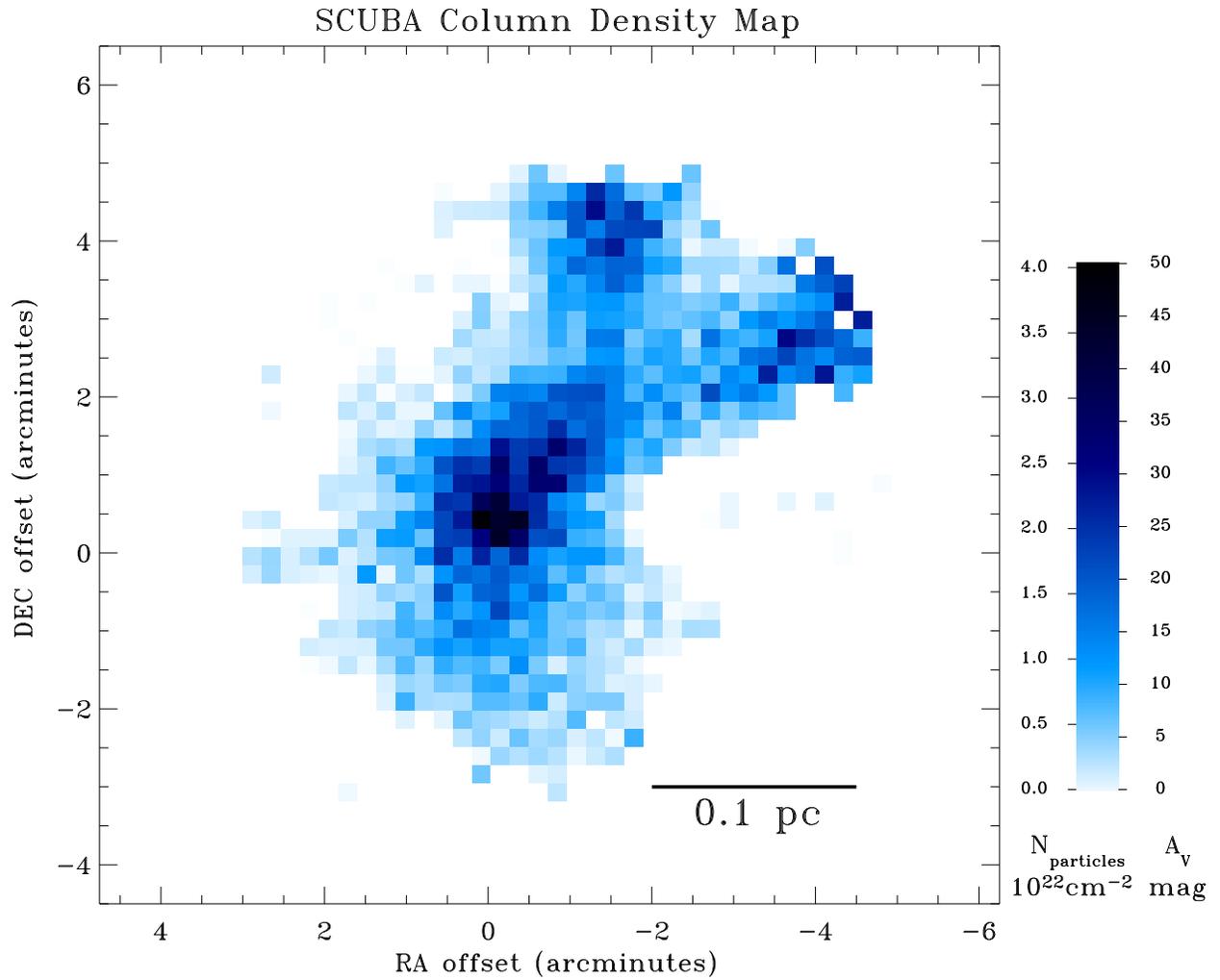}
\caption{Column density derived from the 450 and 850 \micron\ SCUBA maps
 	with the choice of $\beta = 1.5$. The column density of particles 
        is given by $N_{\rm particles} = N_{\rm H}/2.33$, 
	where $N_{\rm H}$ is derived as in Section \ref{EXTINCTION}.  
	\label{EXTMAP}}
\end{figure}

\begin{figure}
\plotone{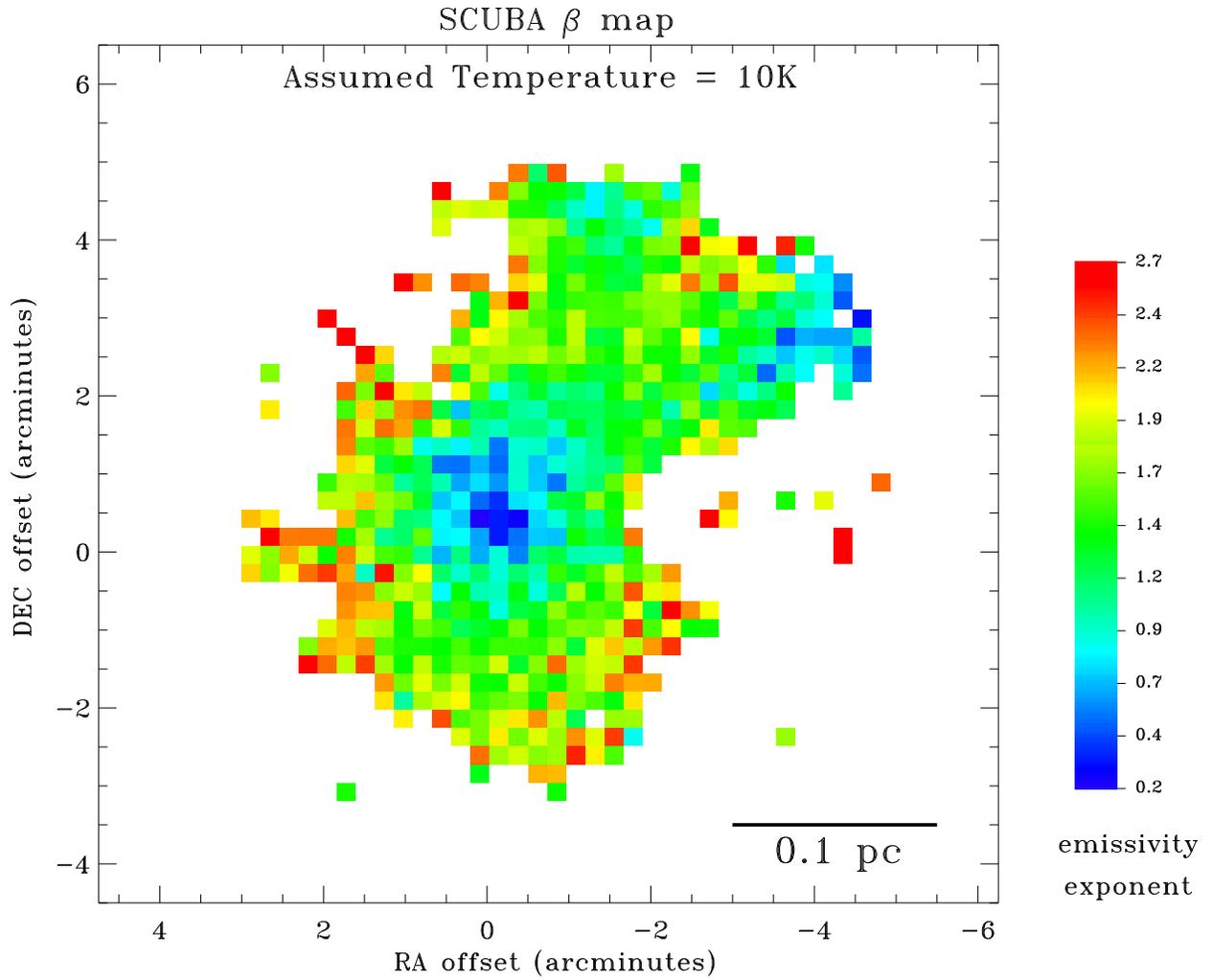}
\caption{Emissivity Spectral Index ($\beta$) of the dust as determined
         by the 450 and 850 \micron\ SCUBA maps with the assumption that the
         dust temperature is constant at 10K. \label{BETAMAP}}
\end{figure}

	The emissivity spectral index, $\beta$, can vary through a
core, though it is often treated as a constant in papers on starless
cores.  By assuming a constant temperature (10 K) and using the 450
\micron\ to 850 \micron\ flux ratio, we have created a map of $\beta$
(Figure \ref{BETAMAP}), in the same way in which the temperature map
\ref{TEMPMAP} was created.  The values of $\beta$ are closer to zero
(i.e. the dust emits more like a blackbody) towards the center of the
core than towards the edges.  This would be expected if the dust
grains are larger at smaller radii in TMC-1C, which might be taken as
evidence of grain growth \citep{Testi03}.  On the other hand, it is
also plausible that the emissivity spectral index does not change much
throughout the core, and instead the temperature of the dust is lower
in the center of the TMC-1C than towards the edges \citep{Zucconi01}.
This would be expected if interstellar radiation is able to heat the
exterior of the core, but the extinction is too high in the center to
allow for efficient heating there.  Note the similarity in the dust
temperature map (Figure \ref{TEMPMAP}) and the emissivity spectral
index map (Figure \ref{BETAMAP}).  The emissivity spectral index
for constant $T = 10$ K has values in the physically plausible range
of 0.5 to 2.0 (center to edge, respectively), making it hard to call
either the constant $\beta$ or constant $T$ assumption unreasonable.
In order to solve for the dust column density, temperature, and
emissivity spectral index we would need a third wavelength in our SED.
   	
\subsection{Mass Estimates With and Without Maps at Multiple Wavelengths} 
	\label{UNIFORM} 
	In most SCUBA observations, non-ideal weather degrades the
quality of 450 \micron\ data significantly more than the 850 \micron\
data.  In such cases, the mass of a core is usually derived from just
850 \micron\ data by assuming that the core is isothermal and assuming
a temperature and emissivity spectral index.  In TMC-1C, because we
can derive a self-consistent temperature map (assuming a value for
$\beta$) from 850 \emph{and} 450 \micron\ data, we can also derive the
mass as a function of radius without assuming a constant temperature.

\begin{figure}
\plotone{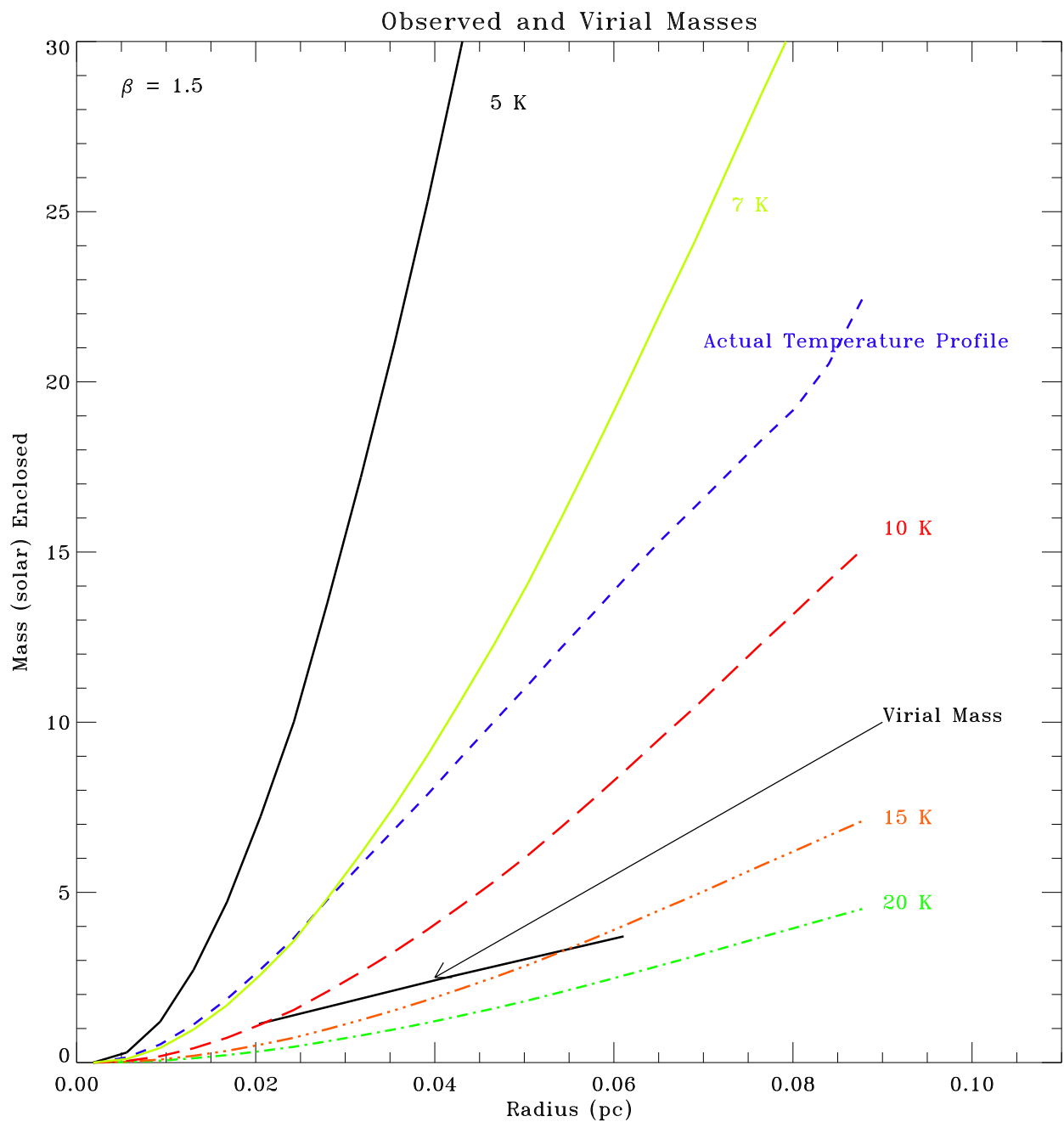}
\caption{The masses plotted here are those contained with a given
         radius of the central sub-mm extinction peak.  The ``virial mass''
         derived from the \nthp\ line width is plotted (assuming a uniform
         density profile and constant temperatuer of 10 K), as is the mass
         derived from the 850 \micron\ SCUBA map coupled with the ``actual
         temperature profile,'' and the masses that would have been derived
         assuming {\it various constant temperatures} and the 850 \micron\ 
	 data alone.  \label{VIRMAP}}
\end{figure}

\begin{figure}
\epsscale{.80}
\plotone{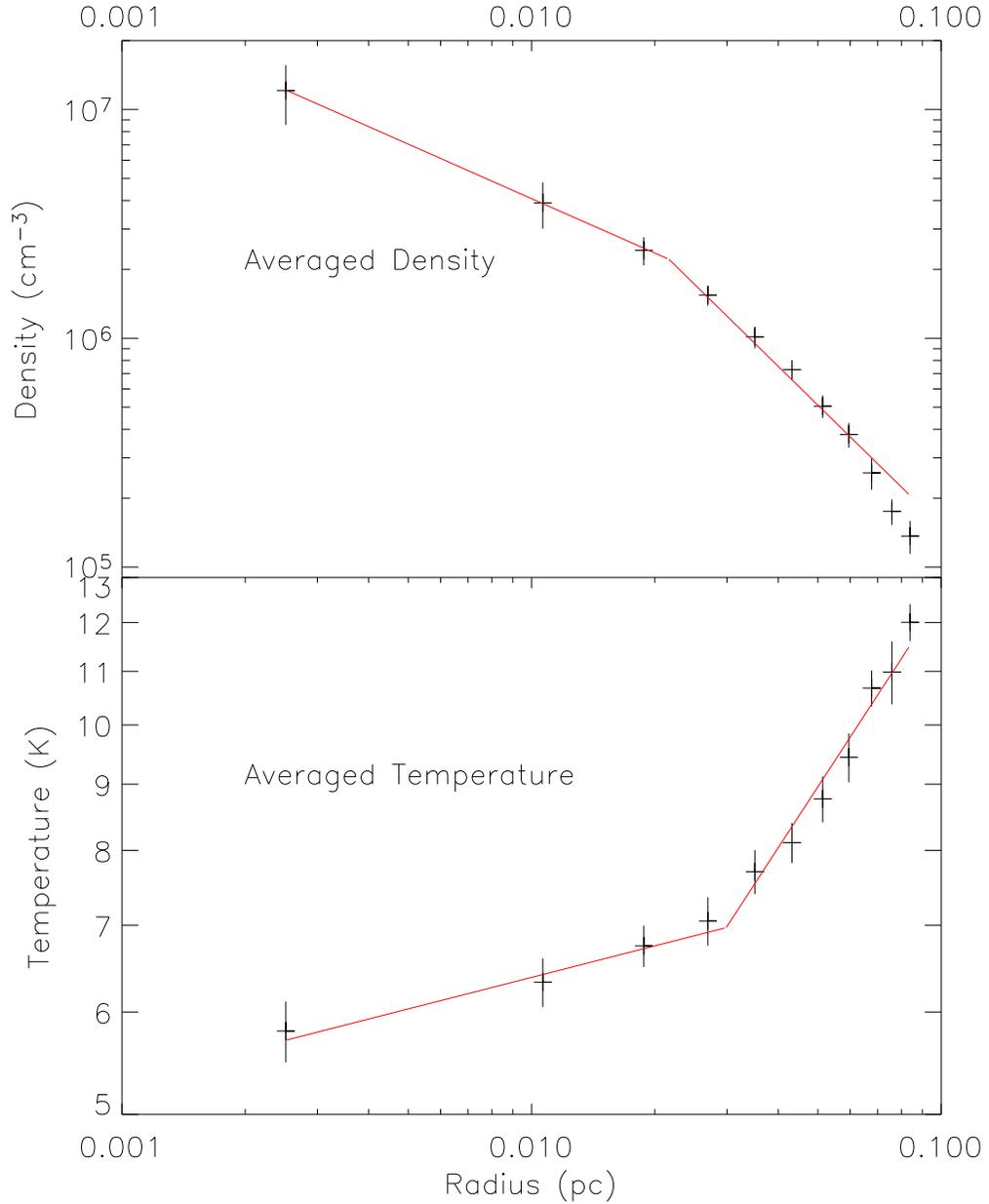}
\caption{These plots show the dependence of density (top) and temperature 
	(bottom) on radius from the center of TMC-1C, which is assumed to be 
	a sphere. The lines show the best fit broken power law, as
        described in Section \ref{PROFILES}.  It is assumed here that 
	$\beta = 1.5$ \label{PROMAPS}}
\end{figure}

\begin{figure}
\epsscale{0.80}
\plotone{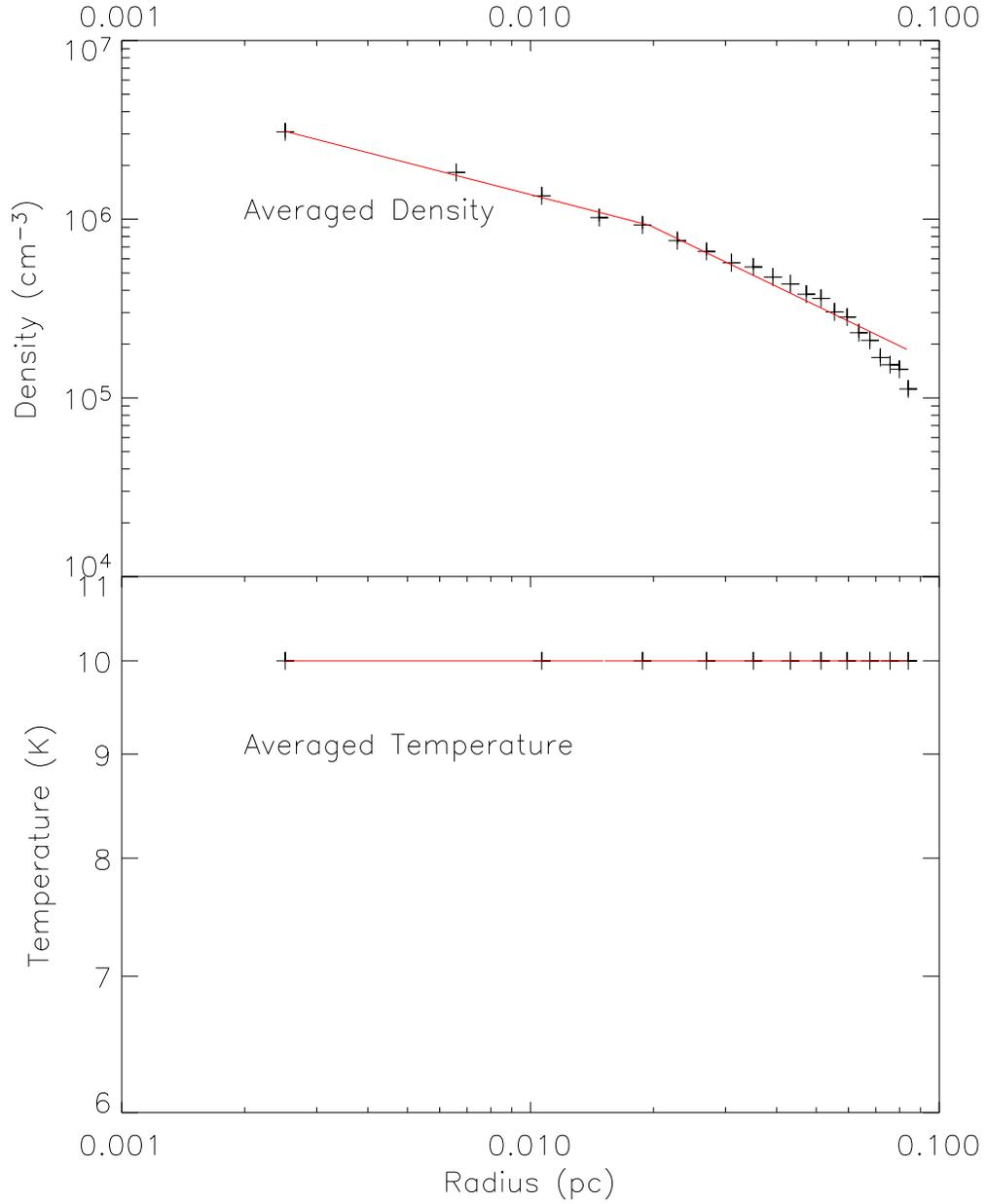}
\caption{These plots show the dependence of density (top) on radius from
         the center of TMC-1C, which is assumed to be an isothermal sphere.
	 The density profile is shown in the top panel for comparison
	 with Figure \ref{PROMAPS}.  The lines show the best fit 
	broken power law, as described in Section \ref{PROFILES}.  It is 
	assumed here that $\beta = 1.5$  \label{CONSTPRO}}
\end{figure}

\begin{figure}
\epsscale{0.80}
\plotone{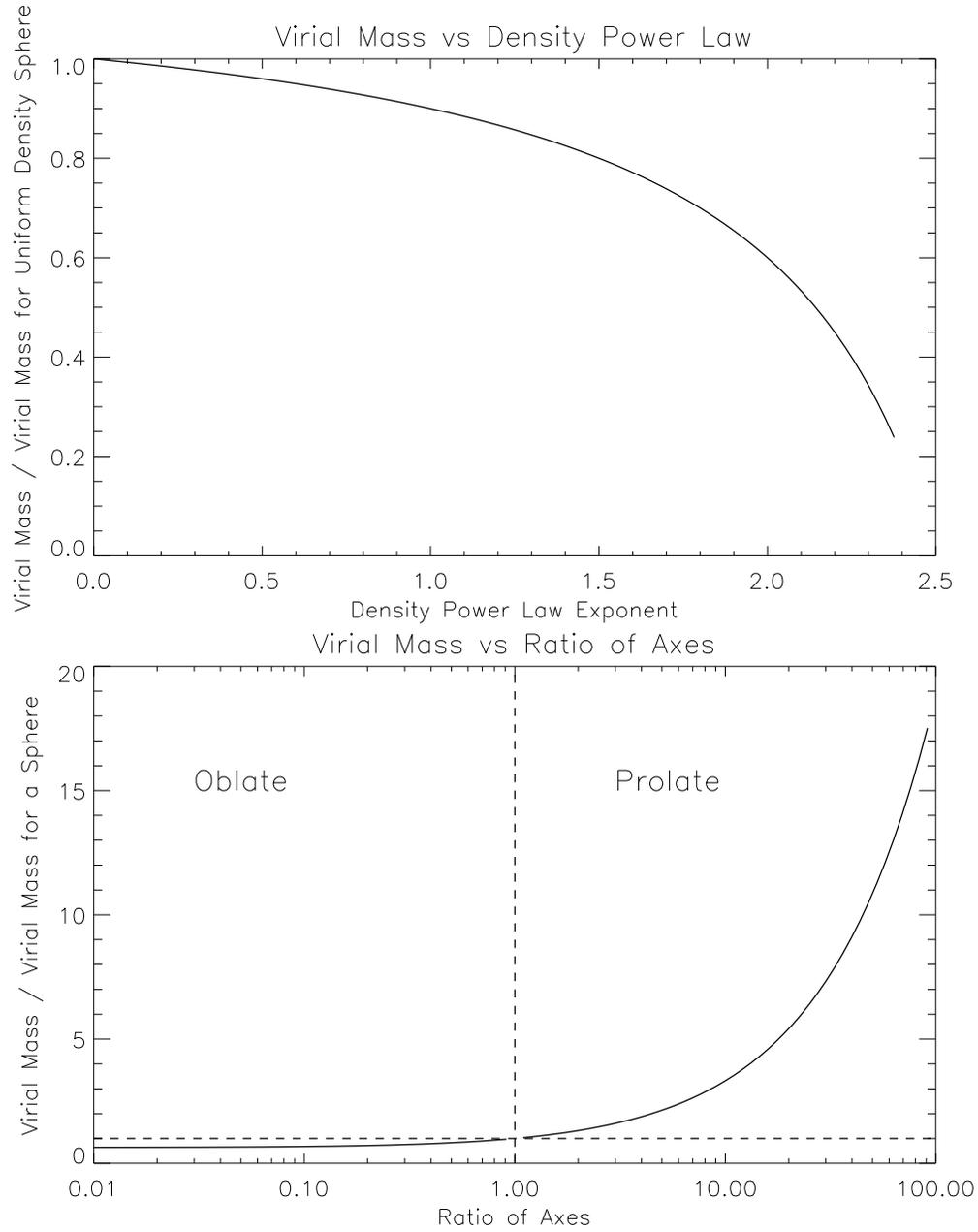}
\caption{(Top) The dependence of the virial mass on the power law exponent
	of the density.  The virial mass is smaller for more centrally
	condensed cores. (Bottom) The dependence of the virial mass on the 
	axial ratio of an elliptical core.  $x < 1$ is an oblate 
	core, and $x > 1$ is a prolate core. Note that raising the virial 
	mass by a factor of 5 requires a prolate shape with an axial ratio 
	of 20:1 (see Section \ref{SHAPEDEP})\label{VIRPARM}}
\end{figure}

\begin{figure}
\epsscale{1.5}
\plotone{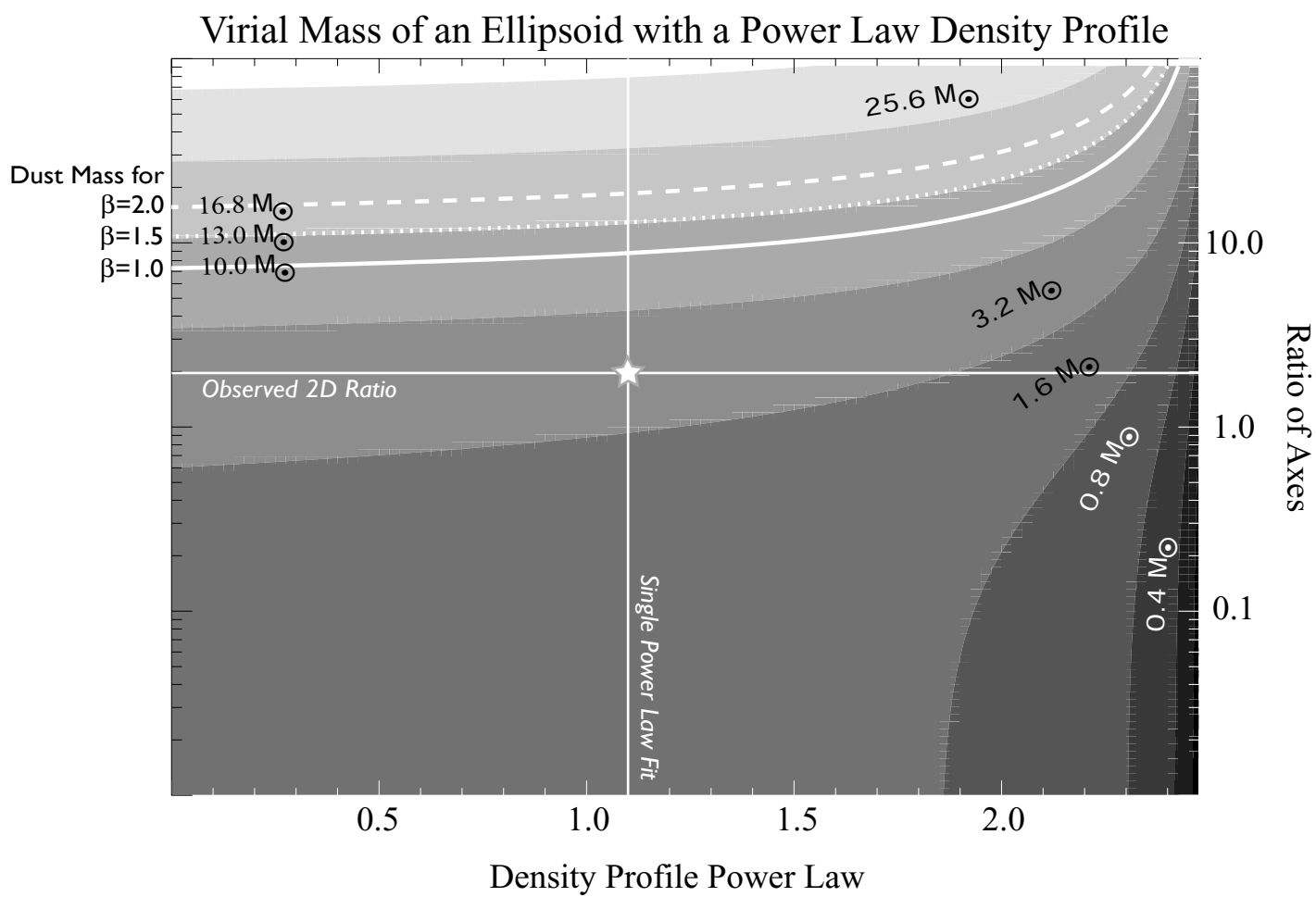}
\caption{The greyscale shows the dependence of the derived virial mass
         of TMC-1C contained within the central 0.06 pc of the core on the 
         ellipticity and density power law exponent.  The virial mass is 
         calculated with the \nthp\ linewidth as in Section \ref{VIRIAL}.
         The highlighted contours show the mass of TMC-1C contained within 
         the central 0.06 pc of the core derived from dust emission with
         $\beta$ taking the values 2.0 (dashed), 1.5 (dotted), and 1.0
         (solid).  The solid horizontal line shows the approximate
         prolate axial ratio of TMC-1C as seen in projection.  The solid
         vertical line shows the best fit single power law density profile
         exponent.  \label{VIRDEP}} 
\end{figure}

\begin{figure}
\epsscale{0.8}
\plotone{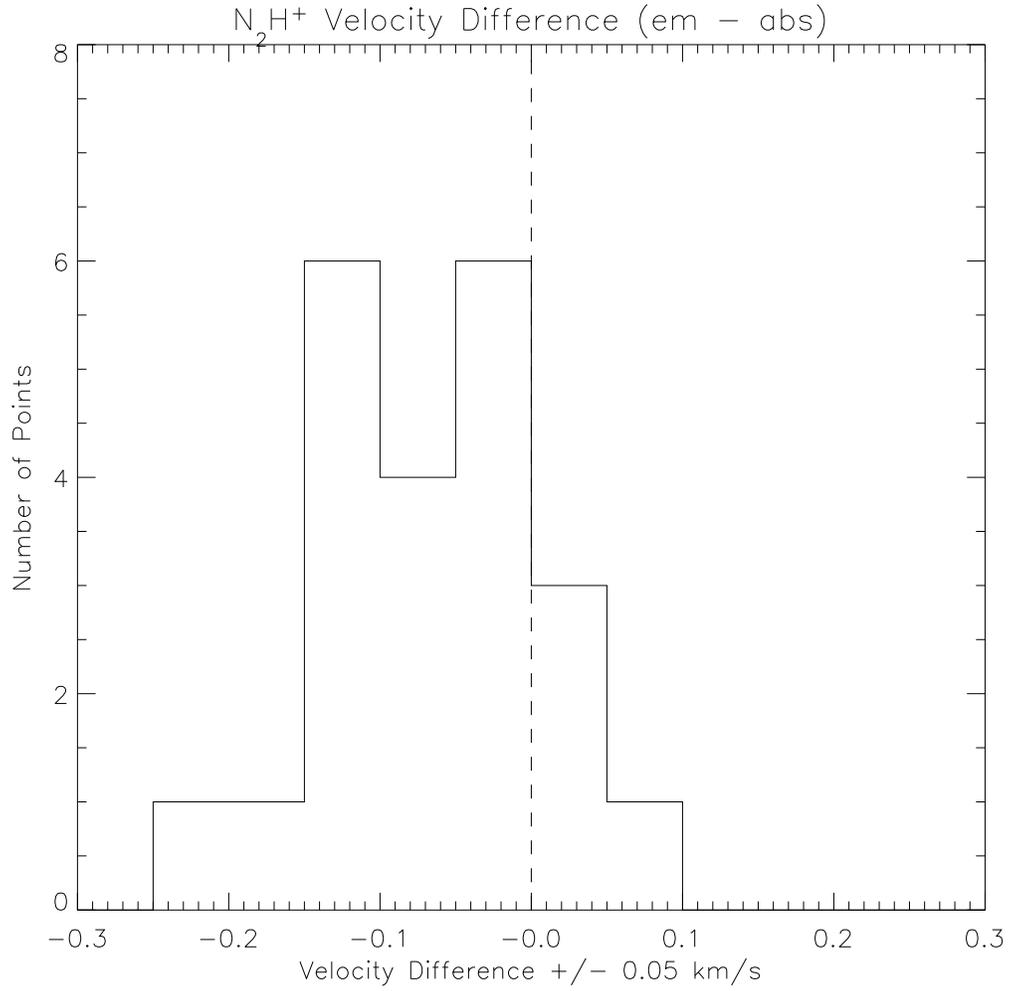}
\caption{The value of the difference between the velocities of the emission
	 and absorption components of the \nthp\ spectra, fit by two 
	 gaussians (see Section \ref{N2H+}). \label{HISTOGRAM}}
\end{figure}

\begin{figure}
\plotone{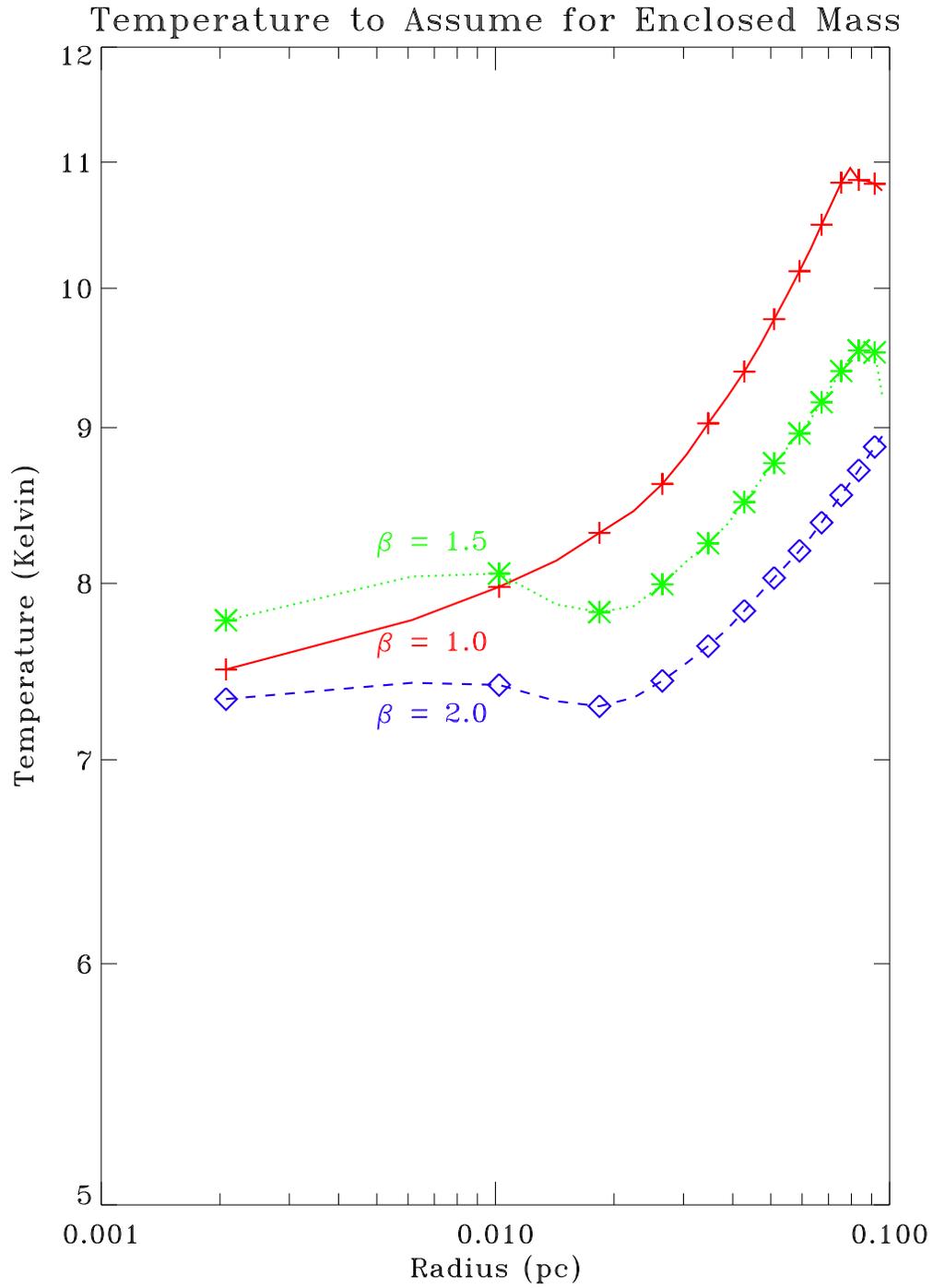}
\caption{This plot shows the temperature that one would need to assume is 
         constant within radius R in order to correctly derive the mass 
         enclosed within that radius from just the total 850 micron flux 
         within that radius for various values of the emissivity spectral 
	 index. \label{ASSUMETEMP}}
\end{figure}

	Here we compare the mass that we derive from our 450 and 850
\micron\ emission maps with what would ordinarily have been derived
from a single wavelength map.  The mass that would have been derived
from just our 850 \micron\ data and a constant temperature is shown in
Figure \ref{VIRMAP} for temperatures of 5, 7, 10, 15, and 20 K.  Note
that the ``true'' cumulative mass profile lies somewhere between the 5
K and 10 K constant temperature mass profiles.  To constrain this
range further, we use the mass profile, M(r), derived from both the
450 and 850 \micron\ fluxes along with the observed 850 \micron\
cumulative flux profile, F(r), to determine a self-consistent
temperature profile, T(r), which is a constant temperature within R.
This profile answers the question ``Given the total 850 \micron\ flux
contained within a cylinder of radius $r$ and the mass contained
within that radius, what isothermal temperature would one need to
assume to make the mass and flux consistent with each other?''  The
result is shown in Figure \ref{ASSUMETEMP}.  The values range from 7 K
to match the inner 0.02 pc of a $\beta = 2.0$ cloud to 11 K to match
the mass out to 0.09 pc of a $\beta = 1.0$ cloud.  This tight range on
the allowed values for a uniform temperature dust model to match the
mass derived from multiwavelength data demonstrates the difficulty of
correctly guessing an appropriate temperature for an isothermal core
and shows the importance of having multiple wavelength bands of
observations to derive it directly.

	The range of isothermal temperatures typically assumed for
cores which do not have multiwavlength data available lies between 30
Kelvin for cores in Orion \citep{Johnstone01, Mitchell01} and
Ophiuchus \citep{Johnstone00} to 10 for cold pre-protostellar cores
\citep{Shirley00}.  Our analysis of TMC-1C shows that the interior
temperature of molecular cloud cores can be significantly colder than
these assumed temperatures, as has been recently shown to be the case
for other cores as well \citep{Evans01}.  The mass of gas and dust in
TMC-1C calculated from the 850 \micron\ flux map for various
isothermal cores is plotted in Figure \ref{VIRMAP}.  If we had assumed
that the dust temperature was uniform at 20 K, then the mass derived
from the dust emission would be \emph{smaller than the virial mass}
for any of the values of $\beta$ that we considered.  A 15 K cloud
would have a dust mass nearly equal to the virial mass for $1.0 \le
\beta \le 2.0$, and a 10 K cloud would have a dust mass considerably
higher than the virial mass for all values of $\beta$.  Because the
inferred gravitational stability of a molecular cloud core is highly
dependent on the assumed temperature, observations should be made at
two or more wavelengths to confidently describe the state of a core.

\subsection{Observed and Model Profiles} \label{COMPARISON}
	Though we have nowhere required that the resulting
temperature, density, or flux profiles of TMC-1C be well fit by
powerlaws, for both the temperature and density profiles, broken power
laws are good fits to the data\footnote{The radius of the innermost
point is taken to be one half of the HWHM of the 850 \micron\ beam.
Though the location of this point is somewhat arbitrary, its influence
on the goodness of fit to the various star formation models is
minimal.}.  In section \ref{N2H+} we found the infall velocity implied
by the \nthp\ spectra, without an attempt to explain the physical
significance of the fit parameters.  In this section we compare the
predictions of several popular analytical star formation models with
the data in order to predict the evolution of the TMC-1C core.  Each
model considered is fit to the density shown in Figure \ref{PROMAPS}.
The fit is done by minimizing the sum of the squared errors between
the data values at each radius, and the predicted density at that
radius.  We have shown that in TMC-1C the inner density profile n(r)
$\propto$ r$^{-0.8}$, the outer density profile is n(r) $\propto$
r$^{-1.8}$, and the infall speed is $\sim$ 0.06 \kms.

	The inside out collapse model of a singular isothermal sphere
\citep{Shu77} predicts a broken power law density distribution with an
r$^{-2}$ profile outside the infall radius, and a profile inside the
infall radius asymptotically approaching r$^{-1.5}$.  Our observations
also show a two-component density distribution, but the inner slope of
the TMC-1C density profile is definitely shallower than predicted by
the Shu model.  To within the errors (see Table \ref{DENSTAB}), the
straight-forward inside-out collapse model can be ruled out by the
density distribution.  In addition, the infall velocity of the inside
out collapse is predicted to be around the sound speed ($a = 0.22$ km
s$^{-1}$ for 10 K gas), while we have observed an infall speed of only
0.06 \kms.

	The logotrope describes a pressure-truncated, self-gravitating
sphere with the equation of state $P/P_c = 1 + A\ln(\rho/\rho_c)$
\citep{McLaughlin96}.  This model predicts a density profile that goes
as r$^{-1}$ in the outer parts of cores and much shallower near the
center while the core is in equilibrium.  Once the cloud begins to
contract, an r$^{-1.5}$ profile develops inside the expansion wave.
The density profile in TMC-1C outer portions of TMC-1C is considerably
steeper than that predicted by the logotropic model, and it is
therefore also inconsistent.

	In many ambipolar diffusion models \citep{Ciolek94,Basu94},
the density of an axisymmetric magnetically regulated cloud can be
described by a broken power law.  In these models, there is an inner
region with a uniform density profile, surrounded by an envelope that
has a power law slope ranging from $r^{-2}$ immediately outside the
uniform density central region and flattening to a mean $\sim$
r$^{-1.4}$ profile.  This qualitatively agrees with our finding for
TMC-1C, in that the inner region of the core has a flatter density
profile than the outer region, though our derived inner powerlaw slope
is significantly shallower than the ambipolar diffusion models
predict.  Certain ambipolar diffusion models developed to match the
conditions (mass, density, rotation, magnetic braking) of the starless
core L1544 do a slightly better job of describing TMC-1C than the
untailored models.  These models predict a steeper profile
($\sim$r$^{-1.8}$) outside the break radius and a flattened interior
profile that match a large portion of the TMC-1C density distribution,
but fail to match the inner density point by wide margin.
\citep{Crutcher94, Safier97}.  In the Crutcher model the infall speed
is as low as 0.033 \kms\ at the boundary of the supercritical core,
and as high as 0.133 \kms\ near the center (7$\times 10^{-4}$ pc).
The infall speed in the central arcminute of TMC-1C, as calculated by
the \nthp\ spectra, is roughly 0.06 \kms, in agreement with this
model.

	The Larson-Penston model describes the uniform collapse of an
isothermal cloud.  Like the expansion wave solution of Shu, the
density in the outer regions of the cloud goes as r$^{-2}$, but it has
a much flatter profile towards the center.  This qualitatively agrees
with the observed TMC-1C density profile.  The infall velocity in the
Larson-Penston model should be $\sim$3.3a, where a is the sound speed
($a = 0.22$ km s$^{-1}$ for 10 K gas).  The TMC-1C infall velocities
are far smaller than the $\sim$ 0.7 km s$^{-1}$ infall velocity
predicted by the Larson-Penston model, and so it too is ruled out.

	The Bonnor-Ebert model describes a pressure bounded isothermal 
sphere, with a density profile that is non-singular at the origin 
\citep{Ebert55, Bonnor56}.  Bonnor-Ebert spheres have a density profile 
that is close to r$^{-2}$ at large radii, and flattens at smaller radii, 
similar to the Larson-Penston model \citep{Harvey01} and qualitatively in 
agreement with the TMC-1C density profile.  The collapse of a Bonnor-Ebert 
sphere has been studied numerically by \citet{Foster93}.  Once the cloud has 
begun to collapse, their model predicts an expanding region of supersonic 
inflow, which asyptotically converges to the Larson-Penston flow.  TMC-1C 
has no evidence of supersonic inflow, so this model fails to describe the 
core, though a Bonnor-Ebert sphere that is just beginning to collapse would 
not necessarily be expected to have supersonic infall motions.  Furthermore, 
the density profile of the Bonnor-Ebert sphere is the best fit to the 
dust-derived density profile for TMC-1C.

	A summary of which models can be ruled out, or not, by their
predicted density profiles and infall velocities is presented in Table
\ref{MODELTAB} and Figure \ref{MASSPROF1}.

\section{Summary}
\subsection{450 and 850 \micron\ data} \label{BEAUTIFUL}
	In sub-mm observations, 450 \micron\ data are often degraded
by weather far more than 850 \micron\ emission.  In this paper we have
shown that obtaining high quality 450 \micron\ data, giving us two
wavelengths near the thermal peak of TMC-1C's modified blackbody
spectrum, is critical to determining the temperature and mass
distribution within a pre-stellar core.
  
\subsection{Low Temperature}
	The temperature map shows that TMC-1C has a strong temperature
gradient if $\beta$ is roughly constant, with the inner regions of the
core being colder than the outer by roughly a factor of two.  This
suggests that no point source has formed in TMC-1C to heat the cloud
from the inside.  The interior temperature of TMC-1C is significantly
colder than 10 K for most values of $\beta$, and rises to $\sim$15
degrees at a radius of 0.1 pc.  This temperature range is lower than
typically assumed for star forming cores, and leads to a much higher
dust and implied gas mass than would have been derived making a
``typical'' assumption of constant $T_{dust}$ of 15 or 20 K.  The
temperatures derived for TMC-1C are consistent with theoretical models
of pre-protostellar cores heated by an external attenuated radiation
field (with or without cosmic rays), which predict cores with a
$\sim$7 K center heating up to $\sim$14 K at the edges
\citep{Zucconi01, Evans01, Galli02}.

\subsection{High Mass} \label{HIGHMASS}
	The virial mass of TMC-1C has been calculated from the line
width of the \nthp\ transition averaged over several annuli.  We show
that the total core mass implied by sub-mm dust emission is
significantly higher than the virial mass for a wide range of density
distributions and axial ratios.  The virial mass at a radius of 0.06
pc is shown in Figure \ref{VIRDEP} as a function of the exponent in
the density power law and ratio of the axes.

	If instead of deriving the temperature from multiwavelength
SCUBA data we had used an assumed temperature of 15 K, this conclusion
would be far less certain.  An assumed constant temperature of 20 K
would have led us to conclude that the virial mass is larger than the
mass determined from dust emission.

\subsection{The Effect of Varying Beta} \label{VARYBETA}	
	For values of the emissivity spectral index in the range from
1 to 2, the mass of the central 0.06 pc of TMC-1C varies from 9
M$_{\odot}$ to 15 M$_{\odot}$, as determined by the same method used
to create Figure \ref{VIRMAP}.  Choosing any value of $\beta$ in this
range does not change the conclusion that the dust derived mass is
much larger than the virial mass.  Furthermore, changing the value of
$\beta$ does not significantly change the shape of the broken power
laws that we fit to the temperature and density profile, and therefore
does not change our fits to various star formation models.  Using
different values of $\beta$ does change the absolute values of the
temperature and density, however, as shown in Figure \ref{PROMAPS}.
If $\beta$ is allowed to vary with radius and if the temperature of
TMC-1C is held constant at 10 K throughout the inner 0.06 pc of the
core, then the value of $\beta$ needs to vary by over a factor of
$\sim$ 3 to account for observations, increasing towards the edges
(see Figure \ref{BETAMAP}).  In reality, it is likely that $\beta$ and
T$_{dust}$ and N$_{dust}$ all vary with radius, but with only two
wavelengths observed, we are forced to hold one quantity fixed to
calculate the other two.  With observations at a third wavelength,
this restriction can be relaxed, and all three quantities ($\beta$,
T$_{dust}$, and N$_{dust}$) can be calculated simultaneously.

\subsection{Infall Motions Detected} \label{INFALL}
	Self-absorbed \nthp\ spectra show evidence of infall motions
of roughly 0.06 \kms\ over a radius of $\sim$0.05 pc.  We believe that
\nthp\ is a good tracer of infall in TMC-1C because it is a high
density tracer, its shape matches the dust emission near the column
density peak, and because it shows no evidence of being significantly
depleted.  We therefore conclude that the interior of TMC-1C is
flowing inwards at about 0.06 \kms.

\subsection{Star Formation Model}
\begin{figure}
\plotone{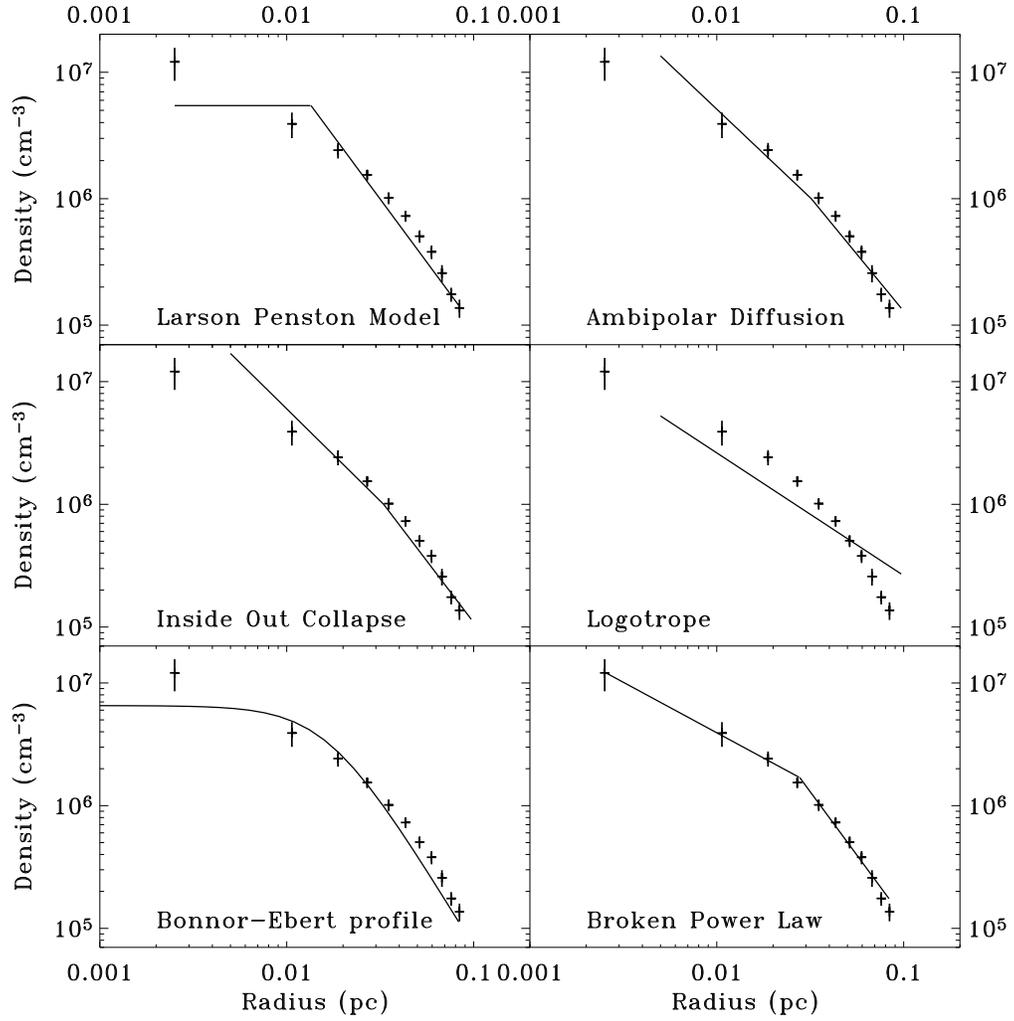}
\caption{Best fit density profiles for TMC-1C, for various models.
         \label{MASSPROF1}}
\end{figure}

	TMC-1C is a starless core that is potentially beginning to
collapse.  Its virial mass is significantly lower than its dust
derived mass, and \nthp\ spectra show signs of sub-sonic infall.  The
derived density profile behaves qualitatively like that predicted by
several star formation models in that it is consistent with an
$r^{-2}$ power law outside of its break radius and is shallower
interior to the break radius.  However, none of the models tested here
convincingly match both the density profile and infall velocity that
we have measured (see Table \ref{MODELTAB} and Figure \ref{MASSPROF1}).

	A Bonnor-Ebert sphere provides the best fit to our
dust-derived density profile, but even for that model the inner point
is only barely within the error bars.  Do our results mean that all
extant detailed models of the collapse of a core into a star are
``wrong''?  Not completely.  The mismatch between theory and
observations here is likely caused both by none of the models, on its
own, being {\it exactly} right, {\it and} remaining assumptions
necessary in the data interpretation being imperfect.  Perhaps
incorporating elements of one theory into another (e.g. ambipolar
diffusion in a Bonnor-Ebert sphere) will produce more realistic
theories.  And, perhaps adding (at least) another wavelength dust map
will modify the $\beta$, N$_{dust}$, and T$_{dust}$ distributions
(mildly), or other spectral lines tracing material even closer to the
dynamical center of cores will reveal (slightly) higher infall speeds.
The excellent weather in Hawaii the day we observed TMC-1C has given
us some of the tightest constraints yet on the physics of isolated
star formation in cores.  Now we need to rise to the challenge offered
by these fine data.

\acknowledgments 
	We would like to thank Paola Caselli, Neal Evans, Eric Keto,
Charlie Lada, Phil Myers, Ramesh Narayan, and Christopher De Vries for
their suggestions, assistance, and insights.  We would also like to
thank Naomi Ridge for assistance with the manuscript.  The James Clerk
Maxwell Telescope is operated by The Joint Astronomy Centre on behalf
of the Particle Physics and Astronomy Research Council of the United
Kingdom, the Netherlands Organisation for Scientific Research, and the
National Research Council of Canada.

\clearpage

\clearpage

\clearpage

\clearpage

\clearpage 

\clearpage

\clearpage

\clearpage 

\clearpage

\clearpage

\clearpage

\clearpage

\clearpage 

\clearpage

\begin{deluxetable}{lrrrrr} 
\tablewidth{0pt}
\tabletypesize{\scriptsize}
\tablecaption{\nthp\ Spectrum at (0,0)\tablenotemark{a} of TMC-1C \label{NTHPTAB}}
\tablehead{
 \colhead{Transition}		& \colhead{Frequency}		&
 \colhead{Signal / Noise}	& \colhead{Noise}		&
 \colhead{V$_{LSR}$}		& \colhead{Line Width}  	\\
 \colhead{}			& \colhead{GHz}			&
 \colhead{}			& \colhead{Kelvin}		&
 \colhead{\kms}			& \colhead{\kms}}
\startdata
 \nthp & 93.176 & 9.8 & 0.145 & -2.787 & 0.246
\enddata
\tablenotetext{a}{(RA,DEC) (J2000) at (0,0) 04:41:38.8 +25:59:42}
\end{deluxetable}
\clearpage
\begin{deluxetable}{lrrrr}
\tablewidth{0pt}
\tabletypesize{\scriptsize}
\tablecaption{Density Profiles of Starless Cores \label{DENSTAB}}
\tablehead{
 \colhead{Observation Type}   	& \colhead{Inner Exponent}	&
 \colhead{Outer Exponent}	& \colhead{Break Radius (pc)}   &
 \colhead{Reference}}
\startdata
 \nthp\ starless cores         & 1.2       & 2         & 0.03  & 1 \\
 L1696 sub-millimeter	       & 1.3       & 2         & 0.02  & 2 \\
 L1689B sub-millimeter         & 1.0 - 1.4 & 2         & 0.02  & 3 \\
 Sub-millimeter starless cores & 1.25      & 2         & 0.02  & 4 \\
 TMC-1C sub-millimeter (3D)    & 0.8 $\pm$ 0.1 & 1.8 $\pm$ 0.1 & 0.02\\
 TMC-1C sub-millimeter (3D, constant $T_{dust})$ & 0.6 & 1.1 & 0.02
\enddata
\tablerefs{
(1) \citep{Caselli02b};
(2) \citep{Ward-Thompson99};
(3) \citep{Andre96};
(4) \citep{Ward-Thompson94}}
\end{deluxetable}
\clearpage
\begin{deluxetable}{lll}
\tablewidth{0pt}
\tabletypesize{\scriptsize}
\tablecaption{Star Formation Models \label{MODELTAB}}
\tablehead{
 \colhead{Model}	& \colhead{Density Profile}	&
 \colhead{Infall Velocity}}
\startdata 
 Bonnor Ebert Sphere	& not ruled out	& not ruled out	\\
 Inside Out Collapse	& ruled out	& ruled out 	\\
 Larson Penston Model	& ruled out	& ruled out	\\
 Logotrope		& ruled out	& not ruled out	\\
 Ambipolar Diffusion	& ruled out     & not ruled out
\enddata
\end{deluxetable}

\end{document}